\documentclass[thmsa,11pt]{article}
\usepackage{amsfonts}

\usepackage{amsmath,amssymb,dsfont}
\usepackage{hyperref}
\usepackage{graphicx}
\usepackage[utf8]{inputenc}
\usepackage{graphicx}
\usepackage{slashed}
\usepackage{color}
\usepackage{bm,bbm}
\usepackage{makecell}
\usepackage[normalem]{ulem}

\usepackage{showlabels}
\usepackage{tikz}

\usepackage{xcolor}%A
\usetikzlibrary{shapes.geometric}

\numberwithin{equation}{section}

\begin{document}

\title{Unconventional SUSY and Conventional Physics\\ 
A Pedagogical Review}

\author{Pedro D. Alvarez$^1$, Lucas Delage$^{2,3}$, Mauricio Valenzuela$^2$ and Jorge Zanelli$^2$ \\
$^1${\small \emph{Departamento de F\'{\i}sica, Universidad de Antofagasta, Aptdo. 02800, Chile}}\\
$^2${\small \emph{Centro de Estudios Cient\'{\i}ficos (CECs), Arturo Prat 514, Valdivia, Chile }}\\
$^3${\small \emph{Instituto de Matem\'atica y F\'{\i}sica, Universidad de Talca, Casilla 747, Talca, Chile.
}}}
\date{}
\maketitle

\begin{abstract}
In supersymmetric extensions of the Standard Model, the observed particles come in fermion-boson pairs necessary for the realization of supersymmetry (SUSY). In spite of the expected abundance of super-partners for all the known particles, not a single supersymmetric pair has been reported to date. Although a hypothetical SUSY breaking mechanism, operating at high energy inaccessible to current experiments cannot be ruled out, this reduces SUSY's predictive power and it is unclear whether SUSY, in its standard form, can help reducing the remaining puzzles of the standard model (SM). Here we argue that SUSY can be realized in a different way, connecting spacetime and internal bosonic symmetries, combining bosonic gauge fields and fermionic matter particles in a single gauge field, a Lie superalgebra-valued connection. In this {\it unconventional} representation, states do not come in SUSY pairs, avoiding the doubling of particles and fields and SUSY is not a fully off-shell invariance of the action. The resulting systems are remarkably simple, closely resembling a standard quantum field theory and SUSY still emerges as a contingent symmetry that depends on the features of the vacuum/ground state. We illustrate the general construction with two examples: i) A 2+1 dimensional system based on the $osp(2,2|2)$ superalgebra, including Lorentz and $u(1)$ generators that describes graphene; ii) A supersymmetric extension of 3+1 conformal gravity with an $SU(2,2|2)$ connection that describes a gauge theory with an emergent chiral symmetry breaking, coupled to gravity. The extensions to higher odd and even dimensions, as well as the extensions to accommodate more general internal symmetries are also outlined.
\end{abstract}


\tableofcontents

\maketitle

%%%%%%%%%%%%%%%%%%%%%%%%%%%%%%%%
\section{Introduction}  %  1  %
%%%%%%%%%%%%%%%%%%%%%%%%%%%%%%%%
The two main frameworks for our current understanding of the fundamental laws of nature, General Relativity ({\bf GR}) and the Standard Model ({\bf SM}) of particle physics, are not only unrelated in scope, but logically independent and seemingly inconsistent with each other. Yet, GR and SM must intersect at some high energy scale in a form that reconciles both schemes. Moreover, in spite of their differences, both frameworks share a common basic feature, {\it Lorentz invariance}.

%%%%%%%%%%%%%%%%%%%%%%%%%%%%%%%%%%%%%%%%%%%%%%%%%%%%
\subsection{Local Poincar\'e symmetry} % 1.1 %
%%%%%%%%%%%%%%%%%%%%%%%%%%%%%%%%%%%%%%%%%%%%%%%%%%%%
GR is a gauge theory for the Lorentz group. This is manifest in the so-called {\it first order formulation}, where the Lorentz connection $\omega^a{}_b$ and the local orthonormal frame (vielbein) $e^a$ are the dynamical variables of the theory, which belong to the adjoint and fundamental representations of the Lorentz group, respectively. The three and four-dimensional Einstein-Hilbert ({\bf EH}) actions --or their Lovelock generalizations to higher dimensions-- are invariant under local Lorentz transformations. The invariance under local translations that would complete local Poincar\'e invariance is not so straightforward, although it can be implemented in a particular class Chern-Simons ({\bf CS}) theories in odd dimensions $D\geq 3$ \cite{HZ}. 

The SM on the other hand, describes elementary particles as irreducible representations of the Poincar\'e group, labeled by mass and spin/helicity Casimir operators of $ISO(3,1)$. In this case, the symmetry is not a gauge invariance, but rather a rigid symmetry of the entire flat spacetime, $\mathbb{R}^{3,1}$, which is an expected symmetry of the in and out states of a scattering process, say. Thus, the internal invariance under local $U(1) \times SU(2) \times SU(3)$ transformations has a very different status from Poincar\'e invariance in the SM. The internal gauge symmetry is responsible for the fundamental interactions among elementary particles --the three fundamental forces of the quantum world--, while the rigid Poincar\'e symmetry classifies the particle states and is not associated to a fundamental interaction in the SM.

The attempt to promote Poincar\'e invariance into a gauge symmetry gives mixed results. As mentioned above, the local Lorentz symmetry can be associated with an interaction field --the Lorentz connection one-form $\omega^a{}_b$, while the translation subgroup of the Poincar\'e group is not well represented by a gauge connection. The closest association of the local translation symmetry would be with the vielbein one-form $e^a$, but there is no four dimensional action $I[\omega, e]$ invariant under $\delta e^a = D\lambda^a$, where $\lambda^a(x)$ is a translation parameter and $D$ is a covariant exterior derivative. Applying such a transformation to the EH action changes it by a bulk term proportional to the torsion, which can only vanish as a consequence of the field equations (as already mentioned, the only exception is the special CS theory invariant under local Poincar\'e transformations, for $D=2n+1\geq 3$). Hence, invariance under local translations would be a {\it contingent symmetry} at best, valid on certain backgrounds satisfying classical torsion constraints. In contrast with this, gauge symmetries are not contingent, they are true symmetries both at the classical and quantum levels. The action of a gauge theory is expected to be invariant (as Yang-Mills or BF theories), or to change by a surface term (as CS), independently of any features of the background. 

In an alternative approach, the translation part of the Poincar\'e group is  identified with the local diffeomorphisms $x^\mu \to x^\mu+ \xi^\mu(x)$, which is an invariance of any generally covariant theory, like GR. The problem with this is that the generators of the diffeomorphism algebra, the Hamiltonian constraints $\mathcal{H}_{\perp}$ and $\mathcal{H}_i$, do not form a Lie algebra but an open algebra, in which the structure constants are replaced by functions of the dynamical fields \cite{Henneaux}. This is not a standard gauge symmetry with a fiber bundle structure that could be naturally combined with the internal gauge invariance for the other interactions. Moreover, it is not clear what could be the quantum mechanical meaning of a symmetry whose nature depends on its on-shell dynamics.

%%%%%%%%%%%%%%%%%%%%%%%%%%%%%%%%%%%%%%%
\subsection{Conventional SUSY} % 1.2 %
%%%%%%%%%%%%%%%%%%%%%%%%%%%%%%%%%%%%%%%
As is well known, the idea of constructing a field theory in which internal and spacetime symmetries are combined into a single Lie group runs into an obstruction known as the Coleman-Mandula no-go theorem. This theorem roughly states that a Lie group that includes Poincar\'e and internal symmetries, that is an invariance of the S-matrix of a standard field theory, is necessarily a direct sum of both Lie algebras and hence trivial \cite{Coleman-Mandula}. This no-go theorem is circumvented if instead of a Lie algebra one considers a graded algebra \cite{HLS}. This observation gave support to the models that were proposed by the pioneers \cite{AV,WZ}, setting off the supersymmetric revolution that culminated with supergravity and superstring theory.

Supersymmetry is a clever scheme to merge spacetime and internal symmetries at the level of the algebras. The trick consists in introducing fermionic generators in an irreducible representation of the Lorentz group (spin $1/2$), also charged with respect to the internal gauge fields (electromagnetic, isospin, color, etc.). Then, the fermionic generators necessarily have nontrivial commutators with both spacetime and internal generators, providing a bridge connecting both types of symmetries. The early SUSY models tried to achieve a supersymmetric extension of the Poincar\'e group, but as mentioned above, it is not easy to construct a gauge theory with local Poincar\'e symmetry. Hence, the first models only considered rigid SUSY transformations rather than gauge symmetries \cite{AV, WZ}. 

Generically, the idea of SUSY was exploited mostly assuming that bosons and fermions belong to a vector-like representation of the type
\begin{equation}
\mathbb{Q}|B \rangle = |F \rangle\;, \qquad \mathbb{Q}|F \rangle = |B \rangle\;,
\end{equation}
where $\mathbb{Q}$ is the generator of supersymmetry and $B,F$ represent bosonic and fermionic states \cite{WZ, PvN}. The superalgebra is completed by typically demanding that applying twice a supersymmetry transformation on a state should be equivalent to the action of a spacetime transformation on that same state. 

Attempts to construct supersymmetric standard models ({\bf SSMs}) long these lines have not been successful in reproducing the observed features of the SM. The main drawback stems from considering the SUSY extension of the Poincar\'e algebra, which leads to supercharges that commute with the Hamiltonian. This immediately implies that particle states connected by SUSY are necessarily degenerate in energy, which leads to a scheme in which for every bosonic particle there is a fermionic one with the same mass and vice-versa. Additionally, since the Poincar\'e group commutes with the internal gauge algebra $u(1)\oplus su(2)\oplus su(3)$, for each boson of a given mass there should exist a fermion not only with the same mass, but also equal electric charge, flavor and color, and vice-versa. The existence of such SUSY pairs, a signal of most SUSY models not seen in the SM, has prompted the argument that SUSY must be spontaneously broken, but should be restored at a sufficiently high energy. The energy scale for this restoration has been pushed to higher and higher scales by larger and larger particle accelerators. With the latest negative results from the LHC  \cite{CMS} hopes have been reduced for ever observing some evidence of SUSY in the lab.

After several decades of struggling to fit the idea of SUSY into the framework of the SM, it seemed that one had two choices, both of doubtful physical relevance: either a rigid SUSY extension of the Poincar\'e group, or a  SUGRA model in which the SM does not seem to fit. This frustrating scenario led many researchers in the 1980s into other lines of inquiry, specially string theory. Still, the question has remained as to the usefulness of the SUSY idea for the SM. 

%%%%%%%%%%%%%%%%%%%%%%%%%%%%%%%%%%%%%%%
\subsection{Gauge SUSY} % 1.3 %
%%%%%%%%%%%%%%%%%%%%%%%%%%%%%%%%%%%%%%%
While the standard SUSY scheme is certainly a natural way to use a graded Lie algebra, it is not the only option. If one wants to include a gauge potential for a fundamental interaction (e. g., a connection field for $u(1)\oplus su(2) \oplus su(3)$) in a SUSY model, a more appropriate representation would be one in which the gauge potential is part of a superconnection that transforms in the adjoint representation of the graded Lie algebra. This would have the added advantage that the supersymmetry could be implemented as a local symmetry --rather than a rigid one-- from the start. Schematically, these superalgebras would have the structure
\begin{align}\label{directsum}
[\mathbb{J}_A, \mathbb{J}_B] = f^C_{AB}\; \mathbb{J}_C\;, \qquad [\mathbb{T}_K, \mathbb{T}_L] &= S^M_{KL}\; \mathbb{T}_M\;, \qquad [\mathbb{J}_A, \mathbb{T}_K] =0 \;, \\ \label{spinfun}
[\mathbb{J}_A, \mathbb{Q}^\alpha_i] = \frac{1}{2} (\Gamma_A)^\alpha{}_\beta \mathbb{Q}^\beta_i \;&,\qquad [\mathbb{T}_K, \mathbb{Q}^\alpha_i] =  (\tau_K)^j{}_i \mathbb{Q}^\alpha_j\;,\\\label{QQ}
\{\mathbb{Q}^\alpha_i, \overline{\mathbb{Q}}^j_\beta\} = \delta^j_i (\Gamma^A)^\alpha{}_\beta \mathbb{J}_A &+ \delta^\alpha_\beta (\tau^K)^j_i \mathbb{T}_K\;,
\end{align}
where $\mathbb{J}$, $\mathbb{T}$ and $\mathbb{Q}$ are the generators of spacetime, internal and SUSY transformations; $\overline{\mathbb{Q}}$ is the charge or Dirac conjugate of $\mathbb{Q}$, and the indices $A, K$ are raised and lowered with the Cartan-Killing form in the respective algebras. Clearly, without the SUSY generators, the bosonic algebra is trivially the direct sum of the spacetime and internal algebras. Commutators \eqref{spinfun} state that the supercharges are in a spinor and the fundamental representations of the spacetime and internal symmetries, respectively. Finally, \eqref{QQ} shows how SUSY ties together spacetime and internal symmetries.

In principle, this idea can be generalize to any two Lie algebras $\mathcal{G}_1$ and $\mathcal{G}_2$, with generators $\{\mathbb{U}_r\}$ and $\{\mathbb{V}_s\}$, instead of spacetime and internal algebras with generators $\{\mathbb{J}_A\}$ and $\{\mathbb{T}_K\}$ discussed above. This would provide a different form of unification rather than the form $\mathcal{G}_1 \oplus \mathcal{G}_2$, like the one assumed in the standard model. Perhaps a gauge SUSY unifying the $U(1)$, $SU(2)$ and $SU(3)$ symmetries of the SM can be developed along this line.

%%%%%%%%%%%%%%%%%%%%%%%%%%%%%%%%%%%%%%%%%%%%%%%%%%%%%%%%
\subsection{Standard and CS supergravities} % 1.4 %
%%%%%%%%%%%%%%%%%%%%%%%%%%%%%%%%%%%%%%%%%%%%%%%%%%%%%%%%
As noted in the pioneering work of Akulov and Volkov \cite{AV}, the consideration of local SUSY leads naturally to theories involving gravity, later referred to as supergravity ({\bf SUGRA}) \cite{PvN,FV}. Phenomenologically, SUGRA has not done much better than SSMs as a low energy SM. A central assumption in most SUGRA theories is that gravity is described by the EH action, with Poincar\'e symmetry playing an essential role in the flat limit. As mentioned above, while the Lorentz group can be easily associated with a local gauge symmetry, the translation part is troublesome in four dimensions \cite{MM}, and replacing the Poincar\'e group by AdS symmetry presents similar issues \cite{T}. This is true even without supersymmetry. The technical obstacle in the construction of a  Poincar\'e (or AdS)-invariant action is the absence of the necessary invariant tensor in even dimensions \cite{N-R,Zu}.

On the other hand, in dimensions $D=2n+1$, the CS forms for $ISO(D-1.1)$, $SO(D-1,2)$ and $SO(D,1)$ provide suitable actions for locally Poincar\'e (and (A)dS)-invariant gravitation theories \cite{Ch,HZ}. In the past three decades, our group has been working on supersymmetric theories in which the dynamical fields are components of a graded Lie-algebra valued connection, that is, a connection associated to a superalgebra. These theories possess local supersymmetry and, if the superalgebra contains the Lorentz or (A)dS algebra, represented by a Lorentz connection and a vielbein, they also include gravity. The gauge fields associated with the SUSY generators are spin 3/2 fermions (gravitini), while the internal symmetry has standard spin 1 gauge fields \cite{TZ, Cast}. These CS supergravity theories are covariant under general coordinate transformations, they do not exhibit matching SUSY pairs and, although they are topological in a certain sense, they describe propagating degrees of freedom \cite{HZ}. In spite of their simplicity and beauty, these models have two drawbacks:  only exist in odd dimensions and they do not contain ordinary spin 1/2 fermions, like the leptons and quarks of the SM, but only spin 3/2 gravitini.

The first problem is related to the fact that no metric is necessary to write down a gauge invariant theory in odd dimensions (using CS forms). In even dimensions, the spacetime symmetry (Poincar\'e, (A)dS, conformal, etc.) breaks down to Lorentz, which means that some pieces (the translation sector) of the connection are no longer gauge fields whilst ``miraculously" provide a metric structure for the theory. This is also the case if a CS theory is reduced to an even dimension below, either by a KK compactification \cite{HTZ,Torabian}, or by an alternative projection onto a submanifold \cite{AWZ}.

As we will see in the next section, the second drawback can be remedied if the gravitini in the connection are interpreted as composite states that combine the metric structure (the vielbein field $e^a_\mu$) and a spin 1/2 fermion.

%%%%%%%%%%%%%%%%%%%%%%%%%%%%%%%%%%%%%%%%%
\section{ Unconventional SUSY} % 2 %
%%%%%%%%%%%%%%%%%%%%%%%%%%%%%%%%%%%%%%%%%
An important point of contrast between the SUSY and the SM refers to the role fermions and bosons play in those theories. In the SM, besides the Higgs field, all bosons are spin 1 fields that carry gauge interactions and they are connections in the adjoint representation of the gauge group. Fermions, on the other hand, are spin 1/2 fields that generate the currents that source the interactions, the building blocks of matter and they are vectors in the fundamental representation of the gauge group. Bosons are one-forms satisfying second order Maxwell-like field equations, while fermions are zero-forms satisfying first order Dirac-like equations. There seems to be not even an approximate symmetry connecting these two radically distinct classes of objects in the SM. Phenomenologically, there seem to be no cases of fundamental bosonic matter fields acting as sources of interactions, or of fundamental interactions mediated by fermions.

%%%%%%%%%%%%%%%%%%%%%%%%%%%%%%%%%%%%%%%%%%%%%%
\subsection{Supersymmetric connection} % 2.1 %
%%%%%%%%%%%%%%%%%%%%%%%%%%%%%%%%%%%%%%%%%%%%%%
To the best of our knowledge, supersymmetry implemented in a superconnection was first explored in \cite{AT}, where a Chern-Simons supergravity theory in 2+1 dimensions for the superalgebras $OSp(2|p)\times OSp(2|q)$ was proposed. This model was also extended to higher odd dimensions as Chern-Simons theories for superconnections with values in graded Lie algebras containing the Poincar\'e or AdS algebras \cite{TZ}. The idea is to consider a connection of the form
\begin{equation} \label{A0}
\mathbb{A} = A^K \; \mathbb{T}_K + \omega^A \, \mathbb{J}_A + \overline{\mathbb{Q}}_\alpha \, \chi^\alpha + \overline{\chi}_\alpha \, \mathbb{Q}^\alpha\;,
\end{equation}
where $\mathbb{T}_K$ are generators of internal symmetry (e.g., $SU(N)$), $\mathbb{J}_A$ is a generator of spacetime transformations (e.g., Lorentz) and $\mathbb{Q}, \overline{\mathbb{Q}}$ are SUSY generators or supercharges, with
\begin{align}
[\mathbb{T}, \mathbb{J}]=0, \;\; [\mathbb{T}, \mathbb{Q}] & \sim\mathbb{Q}, \;\;[\mathbb{J}, \mathbb{Q}]\sim\mathbb{Q} \\
\{\mathbb{Q}, \overline{\mathbb{Q}} \} & \sim \mathbb{T} + \mathbb{J}\;.
\end{align}
As a consequence of the fact the the supercharges belong to the fundamental representation of the internal group, and to the spinor representation of Lorentz, the fermion field $\chi$ will also belong to those representations

From a phenomenological viewpoint, the field $\mathbb{A}$ has an unappealing aspect: its only fermionic component is a  spin 3/2 field $\chi$, the gravitino, an exotic form of gauge field, whose existence has not been confirmed either directly or indirectly, unlike the usual spin 1/2 leptons and quarks of the SM. 

%%%%%%%%%%%%%%%%%%%%%%%%%%%%%%%%%%%%
\subsection{Matter ansatz}  % 2.2 %
%%%%%%%%%%%%%%%%%%%%%%%%%%%%%%%%%%%%
The unconventional SUSY idea ({\bf u-SUSY}) includes an additional assumption that brings in a standard spin 1/2 fermion into this scheme. We refer to this assumption as {\em the matter ansatz}.

In \cite{AVZ} a three-dimensional model was constructed along the line mentioned above, in which the connection (a graded Lie-algebra valued one form) plays a central role. In this superconnection, the gravitino field ($\chi$) is assumed to be a composite of the vielbein ($e^a$) and a spin 1/2 Dirac spinor ($\psi$), and not an irreducible spin 3/2 field,
\begin{equation} \label{m-a}
dx^\mu\;\chi^\alpha_\mu = dx^\mu\; e^a_\mu (\Gamma_a)^\alpha{}_\beta \psi^\beta,
\end{equation}
where $\Gamma_a$ are the Dirac matrices. This ansatz has important consequences.

Without this assumption --\textit{i.e.} leaving the gravitino as usual--, the theories constructed for this connection in odd-dimensions are those in the family of CS supergravities considered in \cite{TZ} which, in 2+1 dimensions have no propagating degrees of freedom (topological field theories) \cite{BGH}.  The ansatz \eqref{m-a} changes the physical nature of the system. In three dimensions, for example, instead of a non propagating gravitino, a propagating spin 1/2 field appears that satisfies a standard Dirac equation minimally coupled to the spin connection $\omega^a$ and to the internal gauge field $A$. The bosonic fields on the other hand, satisfy standard CS dynamical equations for massless gauge connections. 

The resulting u-SUSY actions are by construction generically invariant under the local internal and Lorentz symmetries and under general coordinate transformations. Thus, those generally covariant theories necessarily include gravitation. The background geometry in vacuum is typically a constant curvature spacetime with torsion. The action is generically {\it not invariant} under SUSY, while maximally symmetric vacua are BPS states, and hence supersymmetric under a reduced SUSY \cite{AVZ, APRSZ, GPZ, AVZ2}.

In what follows, we show how u-SUSY can be implemented in odd an even dimensions. These examples are generalizations of the three-dimensional CS action for an $OSp(2|2)$ connection in \cite{AVZ}, and of the MacDowell-Mansouri model in four-dimensions for $OSp(4|1)$ \cite{MM}.  

%%%%%%%%%%%%%%%%%%%%%%%%%%%%%%%%%%%%%%%%
\section{u-SUSY in odd dimensions} % 3 %
%%%%%%%%%%%%%%%%%%%%%%%%%%%%%%%%%%%%%%%%
Let us review first u-SUSY in odd dimensions as this problem is considerably simpler than its even-dimensional counterpart. The simplicity is due to the fact that in order to construct a gauge invariant theory in odd dimensions all that is needed is a connection field, whereas in even dimensions a connection alone is not sufficient and one needs in addition a metric structure. 

%%%%%%%%%%%%%%%%%%%%%%%%%%%%%%%%%%%%%%%%%%%%%%%%
\subsection{Three-dimensional u-SUSYs}   % 3.1 %
%%%%%%%%%%%%%%%%%%%%%%%%%%%%%%%%%%%%%%%%%%%%%%%%
We want to construct an action in three dimensions invariant under local Lorentz and $U(1)$ gauge transformations with these two symmetries connected by SUSY. This could describe a system consisting of an electrically charged spin 1/2 Dirac field interacting with three-dimensional electromagnetic and gravitational fields. 

The smallest supersymmetric Lie algebra containing $so(2,1) \times u(1)$ is $osp(2,2|1)$ has Lorentz generators $\mathbb{J}_a =\frac{1}{2}\epsilon_a{}^{bc}\, \mathbb{J}_{bc}$, $U(1)$ generator $\mathbb{T}$, and supercharges  $\mathbb{Q}^\alpha$ and $\overline{\mathbb{Q}}_\alpha$ ($a,b,c =1,2,3$ and $\alpha =1,2$). The non vanishing (anti-)commutators are
\begin{eqnarray} \nonumber
\; [\mathbb{J}_a,\, \mathbb{J}_b] = \epsilon_{ab}{}^c\, \mathbb{J}_c\,, \qquad \qquad \qquad  \{ \mathbb{Q}^\alpha, \overline{\mathbb{Q}}_\beta \} =-(\Gamma^a)^\alpha{}_\beta \mathbb{J}_a -i \delta^\alpha{}_\beta \mathbb{T}\,, \\  \label{s-algebra} [\mathbb{J}_a, \mathbb{Q}^\alpha] = 1/2 (\Gamma_a)^\alpha{}_\beta \mathbb{Q}^\beta\,, \quad \qquad \;\; [\mathbb{J}_a, \overline{\mathbb{Q}}_\alpha] = -1/2 \overline{\mathbb{Q}}_\beta (\Gamma_a)^\beta{}_\alpha \,,  \qquad \\ \nonumber
\; \; [\mathbb{T}, \mathbb{Q}^\alpha] = i \mathbb{Q}^\alpha\,, \;  \qquad  \qquad \qquad \qquad [\mathbb{T}, \overline{\mathbb{Q}}_\alpha] =-i \overline{\mathbb{Q}}_\alpha \,. \qquad \qquad \qquad
\end{eqnarray}
The corresponding connection field is the one-form valued in the graded algebra is exactly of the form \eqref{A0}, 
where $A=A_\mu dx^\mu$ is the $U(1)$ gauge field, $\omega^a = (1/2)\epsilon^{abc} \omega_{ab \mu} dx^\mu$ is the Lorentz (spin) connection, and $\chi^\alpha= \chi^\alpha_\mu dx^\mu$ is an anticommuting one-form identified as the gravitino. The corresponding Chern-Simons three-form wedge products of exterior forms are implicit and shall not be displayed),
\begin{equation}
\left\langle \mathbb{A} d\mathbb{A} + \frac{2}{3} \mathbb{A}^3 \right\rangle \;,
\end{equation}
where $\langle \cdots \rangle$ is the invariant (super) trace, defines a SUGRA action in three spacetime dimensions \cite{AT}. The field equations imply that the classical configurations are flat,
\begin{equation}
\mathbb{F} = d\mathbb{A} + \mathbb{A}^2 =0\;,
\end{equation}
that is, the super connection $\mathbb{A}$ is locally flat and can be locally expressed as pure gauge, $\mathbb{A} = U^{-1} d U$, where $U(x)$ is a position-dependent element of $OSp(2,2|1)$. The curvatures of the respective subgroups are given by
\begin{equation}
R^a - 2\overline{\chi}\,\Gamma^a\,\chi=0 \,, \, \;\; F-\overline{\chi}\,\Gamma^a\,\chi =0\,, \, \;\; \nabla\chi^\alpha = d\chi^\alpha + iA\chi^\alpha + \omega^a (\Gamma_a)^\alpha{}_\beta \chi^\beta =0.
\end{equation}
where $R^a= d\omega^a + \epsilon^a{}_{bc} \omega^b \omega^c$ and $F=dA$. This theory is a three-dimensional CS SUGRA with no propagating degrees of freedom (CS dynamics) and there is trivially a matching between bosonic and fermionic degrees of freedom (0), an accidental consequence of the topological nature of the theory. 

Assuming the gravitino to be a composite of a spin-1/2 fermion and the vielbein field, as in \eqref{m-a} promotes the fermion $\psi$ into a charged Dirac particle propagating on a curved manifold whose metric structure is defined by the dreibein \cite{GPZ}. In fact, the field equation for $\psi$ is
\begin{equation} \label{Dirac}
    \left[\slashed{\partial} + i \slashed{A} -\frac{1}{4} \Gamma^a \slashed{\omega}_a + e^a_\mu T_{a \nu \lambda} \epsilon^{\mu \nu \lambda} \right]\psi =0.
\end{equation}

The background geometry is characterized by constant curvature and torsion, where the cosmological constant is a constant of integration for the dreibein. This in turn implies that the last term involving torsion in \eqref{Dirac} is also proportional to the same constant and fixes the mass of the fermion. The bosonic fields $\omega, A$ remain massless and are ``locally pure gauge" (for details, see \cite{AVZ}).

Thus, this model describes a massive, spinless charged fermion propagating on a background characterized by a spacetime geometry of constant curvature in the presence of an electromagnetic field given by a locally exact connection. Around the Lorentz-flat vacuum ($R^{ab}=0$, $\psi=0=A$), this is the action for graphene in the long wavelength limit \cite{Iorio}, which suggests that this interesting material can add to its many extraordinary features that of being essentially supersymmetric. This theory can also be seen as a boundary theory of a four dimensional AdS supergravity when reduced by the matter ansatz \cite{ACDT,ACDGNTZ}

The vacuum configuration, AdS$_3$ with $\psi=0=A$, is gauge invariant under the entire $OSp(2,2|1)$ symmetry. For other BPS states, such as the $M=0$ or the $M=J$ BTZ black holes, 1/2 or 1/4 of all possible supersymmetries would be preserved \cite{CH}, and for a generic background SUSY would be at most an approximate symmetry. This type of contingent symmetry also occurs for the invariance under translations, which is at best an approximate symmetry in nature.

One could also consider enlarging the internal gauge symmetry to a non-abelian gauge group such as the case of $U(1)\times SU(2)$ studied in \cite{APRSZ}. Also in that case, the fermion has an additional $SU(2)$ charge (spin) and is the only propagating field; the vacuum structure is richer, with different types of globally defined Killing spinors. The bosonic sector of this theory is rather simple, essentially 2+1 gravity plus locally flat $U(1)$ and $SU(2)$ connections. The propagating spinor endowed with a nonabelian charge would probably make this system phenomenologically quite interesting.

%%%%%%%%%%%%%%%%%%%%%%%%%%%%%%%%%%%%%%%%%%%%
\subsection{Higher odd dimensions}   % 3.2 %
%%%%%%%%%%%%%%%%%%%%%%%%%%%%%%%%%%%%%%%%%%%%
Extending this same scheme to higher odd dimensions $D=2n+1$ is fairly straightforward. The Lorentz group $SO(2n,1)$ can be complemented with an internal gauge group $G_0$; the supercharge $\mathbb{Q}^\alpha_a$ would be in a spin-1/2 representation of the Lorentz group and in the fundamental of $G_0$. The connection $\mathbb{A}$  and the corresponding CS forms, with the matter ansatz, are uniquely defined and prescribe a field theory with gauge supersymmetry involving gravity, gauge fields and charged spin 1/2 fermions. The bosonic part of the connection would generically propagate and SUSY would not be generically respected. Nevertheless, the maximally symmetric vacua would still be invariant under the entire supersymmetric group and therefore SUSY would be a contingent symmetry. 

A five-dimensional CS model along these lines was studied in \cite{GHN}. The general problem of a generic CS supergravity theory with the additional assumption of the matter ansatz has not been addressed yet.

%%%%%%%%%%%%%%%%%%%%%%%%%%%%%%%%%%%%%%%%%%%
\section{ u-SUSY in even dimensions}  % 4 %
%%%%%%%%%%%%%%%%%%%%%%%%%%%%%%%%%%%%%%%%%%%
In even dimensions, the previous construction cannot be carried out because there exist no Chern-Simons forms to define the Lagrangian in that case. One can try to build a Yang-Mills or a BF action with a super connection of the form \eqref{A0} for a given superalgebra. This, however has the important consequence that the gauge symmetry must be necessarily broken since there is no local action that is invariant under the entire gauge superalgebra. This was already observed in the pioneering work of MacDowell and Mansouri \cite{MM} and Townsend \cite{PKT} emphasized in a more general context by Regge and collaborators \cite{N-R,Cast,Dauria-Fre-Regge}.

In the end, this situation can be seen as a consequence of the Chern-Weil theorem \cite{Nakahara}. Suppose one wants to write down an action in $2n$ dimensions for a connection $\mathbb{A}$ that transforms in the adjoint representation of a gauge group $G$, $\mathbb{A} \rightarrow U^{-1} (\mathbb{A} + d)U$, where $U\in G$. Then, $\mathbb{F}^{a} \rightarrow U^{-1} \mathbb{F}\,U$, where $\mathbb{F}^{a} = d\mathbb{A}^{a} + f^a_{bc}\mathbb{A}^{b} \mathbb{A}^{c}$, and any functional of the form
\begin{equation} \label{CW}
I[\mathbb{A}]= \int C_{a_1 a_2 \cdots a_n} \mathbb{F}^{a_1} \mathbb{F}^{a_2} \cdots \mathbb{F}^{a_n} \;, 
\end{equation}
that is invariant under gauge transformations, is a characteristic class and therefore not a useful action. Consequently, two possibilities arise for a nontrivial theory invariant under a gauge symmetry group $G$: \\
1) The fields in the action cannot be only the connection $\mathbb{A}$ in $\mathcal{G}$, the Lie algebra of $G$. Besides the connection, other fields must enter in the Lagrangian. These other fields must combine in a way to make a $G$-invariant action, or\\
2) The Lagrangian is constructed with the $\mathcal{G}$-valued connection $\mathbb{A}$, but the action is not invariant under the entire group $G$. The symmetry is broken down to a subgroup $G' \subseteq G$ because $C_{a_1 a_2 \cdots a_n}$ is not an invariant tensor of $G$ but an invariant of $G'$. Then, the effective gauge symmetry of the action is $G'$ and the part of the connection that carries the generators of $G'$ is the only true gauge connection, while the part corresponding to the broken symmetries are in a vector representation of $G'$.

%%%%%%%%%%%%%%%%%%%%%%%%%%%%%%%%%%%%%%%%
\subsection{4D gauge gravities}  % 4.1 %
%%%%%%%%%%%%%%%%%%%%%%%%%%%%%%%%%%%%%%%%
An example of the first situation is the Einstein-Hilbert action in 4 dimensions,
\begin{equation}\label{EH}
I[\omega, e]= \int_{M^4}\epsilon_{abcd}\left(R^{ab} + \alpha e^a e^b \right)e^c e^d \;,
\end{equation}
where $R^{ab}=d\omega^{ab} + \omega^a{}_c\;\omega^c{}_b$ is the curvature and $\omega$ is the connection for the Lorentz group $SO(3,1)$, while $e^a$ {\it is not a connection} but a Lorentz vector and enters in the Lagrangian is such a combination that makes the expression \eqref{EH} invariant under $SO(3,1)$.

One can also understand this situation from the second viewpoint. Consider combining $\omega^a{}_b$ and  $e^a$ into a connection for a semisimple group $G\supseteq SO(3,1)$ (take $G$ as $SO(3,2)$ or $SO(4,1)$),
\begin{equation} \label{W}
\mathbb{W}^{AB} = \left[ \begin{array}{cc}
  \omega^{ab} & e^a/\ell \\
  -e^b/\ell   & 0
\end{array}
\right] \;.
\end{equation}
It is necessary to introduce the constant $\ell$ with units of length in order to match the dimensionless nature of the connection $\omega$. Here the upper case indices take values $0,1,2,3,4$. Then, the curvature $\mathbb{F}^{AB}=d \mathbb{W}^{AB} + \mathbb{W}^A{}_C\;\mathbb{W}^{CB}$ is
\begin{equation} \label{F}
\mathbb{F}^{AB} = \left[ \begin{array}{cc}
  R^{ab} \pm e^a\,e^b/\ell^2  & T^a/\ell \\
  -T^b/\ell   & 0
\end{array}
\right] \;,
\end{equation}
which can be used to construct a Lagrangian for a four-dimensional action (the $\pm$ signs correspond to AdS ($+$) or dS ($-$)),
\begin{equation} \label{CW-F}
I[\mathbb{W}]= \int C_{ABCD} \mathbb{F}^{AB} \mathbb{F}^{CD} \;.  
\end{equation}

The problem is, as MacDowell and Mansouri noted, that there exist no $G$-invariant fourth rank tensors to be used as $C_{ABCD}$. The only invariant fourth rank tensors with the right symmetry are 
\begin{equation}
\epsilon_{ABCD} \qquad \mbox{or} \qquad [\eta_{AC} \eta_{BD} -\eta_{AD} \eta_{BC}].
\end{equation}
Unfortunately, the fourth rank Levi-Civita tensor is only an invariant for $SO(3,1)\subseteq G$, and therefore it is not possible to write \eqref{CW-F} with $C=\epsilon$. Taking the second invariant tensor in \eqref{CW-F} yields another characteristic class, the Pontryagin invariant for $G$,
\begin{align} \label{P-G}
I_{P}[\mathbb{W}] & = \int \mathbb{F}^{AB} \mathbb{F}_{AB} \;,
\end{align}

Alternatively,using the curvature for $G$ but with $\epsilon_{abcd}$, the invariant tensor for $SO(3,1)$, yields the Einstein-Hilbert action (plus the Euler characteristic),
\begin{align} \nonumber
I_{EH}[\mathbb{W}] & = \int \epsilon_{abcd} \mathbb{F}^{ab} \mathbb{F}^{cd} \\ \label{EH'}
& = \int \epsilon_{abcd} \left(R^{ab} \pm e^a e^b/\ell^2 \right) \left(R^{cd} \pm e^c e^d/\ell^2 \right) 
\end{align}
Alternatively, using the curvature $\mathbb{F}$ and the other $SO(3,1)$ invariant, $[\eta_{ac} \eta_{bd} -\eta_{ad} \eta_{bc}]/2$ yields
\begin{align} \nonumber
I_{T}[\mathbb{W}] & = \int \left(R^{ab} \pm e^a e^b/\ell^2 \right) \left(R_{ab} \pm e_a e_b/\ell^2 \right)  \\ \label{Tor}
& = \int \left(R^{ab}R_{ab} \pm 2R^{ab}e_a e_b/\ell^2 \right)\;,
\end{align}
which, apart from the Lorentz-Pontryagin invariant $R^{ab}R_{ab}$, is a somewhat exotic but legitimate four-dimensional action for a theory with torsion, but it is only invariant under $SO(3,1) \subseteq G$.

In the end, both scenarios (1) and (2) above, lead to the same conclusion: in $4$ dimensions a local action invariant under $SO(4-p,p)$ must contain some fields besides an $so(4-p,p)$ connection. There is still a consolation prize for the theories that are not invariant under the full $G$ symmetry. The backgrounds defined by $\mathbb{F}^{AB} =0$ are gauge invariant under full local $G$ transformations. (This is rather trivial since $\mathbb{F} =0 \Rightarrow $ $U^{-1}\,\mathbb{F}\,U = 0$.)

%%%%%%%%%%%%%%%%%%%%%%%%%%%%%%%%%%%%%%%%%%%%%%%%%%%%%%%%%%%%
\subsection{Yang-Mills action and generalizations}   % 4.2 %
%%%%%%%%%%%%%%%%%%%%%%%%%%%%%%%%%%%%%%%%%%%%%%%%%%%%%%%%%%%%
One could object to the previous discussion arguing that the Yang-Mills ({\bf YM}) gauge action 
\begin{equation} \label{YM}
I_{YM}[\mathbb{A}] = \int_M Tr\left[\mathbb{F}\;  \ast \mathbb{F}\right]\;,
\end{equation}
built for the connection in the Lie algebra of the group $G$ is invariant under same group in any dimension. 

The point is that the Hodge dual $^*\mathbb{F}$ is a function of $\mathbb{F}$ and of the {\it metric structure} of $M$. This may seem like a small detail but it is a relevant one. Even if in some manifolds and in special coordinates the metric can taken a rather simple form, such as $g_{\mu \nu} = diag(-1,+1,+1,+1)$, the metric is a field and it changes under deformations of $M$ (and under changes of coordinates). This sensitivity of the metric to the changes in $M$ implies that $I_{YM}$ is not a topological invariant.

Similarly, the integral of the invariant $2n$-form $Tr[\mathbb{F}^n]$ over a compact manifold is a topological invariant, but if multiplied by a scalar field $\phi(x)$, then the functional 
\begin{equation} \label{phi-T}
I[\mathbb{A},\phi] = \int_M \phi \; Tr\left[\mathbb{F}^n\right]\;,
\end{equation}
is not a topological invariant. It defines a sensible action, with reasonable equations of motion.

%%%%%%%%%%%%%%%%%%%%%%%%%%%%%%%%%%%%%%%%%%%%%%%%
\subsection{Four-dimensional u-SUSY}   % 4.3 %
%%%%%%%%%%%%%%%%%%%%%%%%%%%%%%%%%%%%%%%%%%%%%%%%
Consider one of the simplest nontrivial superalgebra in 4 dimensions containing Lorentz and an abelian internal gauge symmetries, $\mathcal{G} \supset so(3,1) \oplus u(1)$. The supercharges are complex Lorentz spinors and  a cursory check shows that the closure of the superalgebra requires the introduction of additional vector generators in the spacetime algebra. The resulting superalgebra is the $osp(4|2) \sim usp(2, 2|1)$, which includes the (A)dS$_4$ generators $\{\mathbb{J}_a\}$ and $\{\mathbb{J}_{ab}\}$, the complex supercharge $\mathbb{Q}^\alpha, \overline{\mathbb{Q}}_\beta$ in a spin 1/2 representation, and the $u(1)$ generator $\mathbb{K}$. The essential anticommutator in the superalgebra reads
\begin{equation} 
\{\mathbb{Q}^\alpha, \overline{\mathbb{Q}}_\beta \} =-\frac{1}{2}(\Gamma^{ab})^\alpha{}_\beta \mathbb{J}_{ab} +(\Gamma^a)^\alpha{}_\beta \mathbb{J}_a -i \delta^\alpha{}_\beta \mathbb{K}\, , 
\end{equation}
where $\Gamma^{ab}=[\Gamma^a,\Gamma^b]/2$ and $\{\Gamma^a\}$ are the Dirac matrices. Adopting the matter ansatz \eqref{m-a}, the connection \eqref{A0} is 
\begin{equation} \label{A1}
\mathbb{A} = A\, \mathbb{K} + f^a\, \mathbb{J}_a\; +\frac{1}{2}\omega^{ab} \, \mathbb{J}_{ab} +  \overline{\mathbb{Q}}_\alpha\,(\slashed{e})^\alpha{}_\beta \psi^\beta + \overline{\psi}_\alpha \,(\slashed{e})^\alpha{}_\beta  \mathbb{Q}^\beta\;,
\end{equation}
where $A,\, f,\, \omega,\, \slashed{e}=\Gamma_a \;e^a$ are one-form fields. The corresponding curvature two-form reads $\mathbb{F} = F_0\,\mathbb{K} + F^a\, \mathbb{J}_a\; + \frac{1}{2}F^{ab}\,\mathbb{J}_{ab} +  \overline{\mathbb{Q}}_\alpha\, \mathcal{F}^\alpha + \overline{\mathcal{F}}_\alpha\, \mathbb{Q}^\alpha$, where
\begin{align}
F_0 &= dA - \overline{\psi} \slashed{e} \slashed{e} \psi \\
F^a &= Df^a + i \overline{\psi} \slashed{e}\; \Gamma^a\; \slashed{e} \psi \\
F^{ab} &= R^{ab} + f^a\; f^b + i \overline{\psi} \slashed{e}\; \Gamma^{ab}\; \slashed{e} \psi \\
\mathcal{F} &= \nabla(\slashed{e} \psi) \\
\overline{\mathcal{F}} &= (\overline{\psi} \; \slashed{e}) \overleftarrow{\nabla}\,,
\end{align}
with $Df^a =df^a + \omega^a{}_b f^b$ and $R^a{}_b = d\omega^a{}_b + \omega^a{}_c \omega^c{}_b$. Additionally, $\nabla = [d - iA + \frac{1}{2} \slashed{f} + \frac{1}{2} \slashed{\omega}]$ is the covariant derivative in the spin-1/2 representation, and  
$[-\overleftarrow{d}- iA + \frac{1}{2} \slashed{f} + \frac{1}{2} \slashed{\omega}] = \overleftarrow{\nabla}$. We observe that the one-form $f^a=f^a_\mu dx^\mu$ has the same features as the metric structure $e^a$, except for the dimensionless nature of the connection and the dimension of length of vielbein.

We know that the 4-form $\langle \mathbb{F}\, \mathbb{F}\rangle$, where $\langle \cdots \rangle$ is an invariant trace in the Lie superalgebra defines a characteristic class, not a Lagrangian. In order to build an action, we need to define some operation $\circledast$ that takes the role of the Hodge-$*$ in YM theories. 
This must be an automorphism in the algebra operation with the property $\circledast^2 =-1$. Moreover, we expect that in the purely bosonic sector the Lagrangian reduces to the Einstein-Maxwell theory. We observe that $\circledast$ should act on the $u(1)$ subalgebra as $\circledast F_0 = *F_0$; on the fermionic sector as $\circledast \mathcal{F} = \Gamma_5 \mathcal{F}$; and on the spacetime subalgebra as  $\circledast F^{AB} = \mathbb{J}_5 F^{AB}$, where $\mathbb{J}_5 \equiv \frac{1}{4!} \epsilon^{abcd} \mathbb{J}_{ab} \mathbb{J}_{cd}$. The definition of $\circledast$ is not unique but this choice seems to be the simplest option and yields the Einstein-Maxwell-Dirac Lagrangian \cite{AVZ2, ADVZ, APZ, AVZ3}.

Note that as in the discussion of $\S$4.1, the spacetime symmetry is broken from $SO(3,2)$ down to $SO(3,1)$. This means that the field $f^a$ is no longer a gauge connection but an ordinary Lorentz vector one-form, indistinguishable from the vielbein, except for the dimensions. Hence, we have two possible metric structures, one defined by $e^a_\mu$, other defined by $\ell f^a_\mu$ for some scale factor $\ell$ with units of length. Another way to look at this problem is to observe that it is not necessary to include a vielbein, because the connection already has a field suited for the job of providing a metric structure. All that is needed is the parameter $\ell$ that sets the length scale of the system.

Putting all these ingredients together, one finds the rather standard looking Lagrangian (see \cite{APZ, AVZ3} for details),
\begin{align} \nonumber
L = &-\frac{1}{4} \langle \mathbb{F}\,\circledast\; \mathbb{F}\rangle \\ \nonumber
= &-\frac{1}{4}F_{\mu \nu}\,F^{\mu \nu} |e|d^4x -\frac{1}{16}\epsilon_{abcd} (R^{ab} + e^a e^b/\ell^2) (R^{cd} + e^c e^d/\ell^2) \\ \nonumber  &-\frac{i}{2\ell}\left[(\overline{\psi}\overleftarrow{D}) \Gamma^a \psi- \overline{\psi}\Gamma^a (\overrightarrow{D} \psi)\right] \epsilon_{abdc} e^b e^c e^d + \frac{2i}{\ell} \overline{\psi}\Gamma_5\slashed{e} \psi(T_b e^b)\\  \label{L4}
&-\frac{i}{2\ell^2} \overline{\psi} \psi \epsilon_{abdc} e^a e^b e^c e^d + 12 \left[(\overline{\psi} \Gamma_5 \psi)^2 - (\overline{\psi} \psi)^2 \right]|e| d^4x\,,
\end{align}
where we have identified $f^a =\ell^{-1} e^a$. The first two pieces in the right hand side are the Maxwell and Einstein-Hilbert Lagrangians with negative cosmological constant; the third term is the Dirac Lagrangian for a charged spin 1/2 field coupled to gravity; the fourth term is a non-minimal coupling of the fermion to torsion; the fifth is a mass term and the last is a Nambu--Jona-Lasinio fermion self-coupling. This latter coupling, is nonrenormalizable and occurs naturally in BCS superconductivity as a natural mechanism to produce a mass gap and chiral symmetry breaking. This is a combined effect of nonrenormalizability and the logarithmic dependence of the mass gap on the cut-off scale \cite{NJL}.

More realistic scenarios in four dimensions can be --and have been-- considered, where the internal gauge symmetry is not just $U(1)$, but a nonabelian one. In fact, in \cite{AVZ2} the construction is given for a u-SUSY theory for $su(2,2|2)$, which is the supersymmetric completion of the conformal group $SO(4,2)$ and contains $U(1) \times SU(2)$ as internal gauge symmetry. Again, the introduction of the $\circledast$ operation reduces the symmetry of the action down to $SO(3,1)\times SU(2) \times U(1)$. There, a refined version of $\circledast$ allow us to write down the Lagrangian in the form,
\begin{equation}
L_{gen} = -\frac{1}{4} \langle \mathbb{F}\,\circledast\; \mathbb{F}\rangle\label{Lgen}
\end{equation}
where now $\mathbb{F}\,\in \, su(2,2|2)$. The latter model, after identification of the vielbein field with and the translation gauge fields in $su(2,2|2)$ yields an action principle for a Dirac field charged under the $u(1)\oplus su(2)$ internal algebra, plus standard gravity and Yang-Mills sectors, with additional NJL terms and couplings to the torsion.

%%%%%%%%%%%%%%%%%%%%%%%%%%%%%%%%%%%%%%%%%%%%%%%%%%%
\section{Discussion and summary and outlook}  % 5 %
%%%%%%%%%%%%%%%%%%%%%%%%%%%%%%%%%%%%%%%%%%%%%%%%%%%
The u-SUSY idea is based on two logically independent ideas: i) The use of a connection in a super algebra that contains a spacetime subalgebra and an internal one, and ii) The matter ansatz \eqref{m-a}. The first, in odd dimensions leads to the family of CS supergravities and in even dimensions to a new and not completely surveyed family of supergravities with gravitini that are charged with respect to the internal gauge symmetry, whose relation with standard  SUGRAs is not completely understood \cite{ADVZ}. The second idea could also be applied to any of the known SUGRAs and it is expected that it would yield a theory containing spin 1/2 fermions in a Dirac-like action coupled to a host of other gauge and matter fields.

We have argued that the inclusion of gravity in high energy models similar to the Standard Model can be quite natural in u-SUSY. In fact, gravitation is essentially unavoidable in u-SUSY because the Lorentz algebra is required to close the superalgebra, as for instance in \eqref{QQ}. In addition, since the matter ansatz requires a metric structure, this completes the list of ingredients for gravity. This type of SUSY matches the observed patterns of massless bosonic carriers of gauge interactions and spin 1/2 (possibly massive) fermions charged with respect to those interactions, and not the other way around. The resulting models don't exhibit supersymmetric pairs, with supersymmetry emerging as an approximate invariance but generically broken, a contingent symmetry like translation invariance. This SUSY breaking is not of the spontaneous kind. The symmetry is not broken by the vacuum but by the action. In fact, the vacuum has a larger symmetry than the action, in a similar way to what happens in Einstein-Hilbert gravity, where the action is not invariant under local Poincar\'e or AdS transformations, but the vacuum is. In u-SUSY, supersymmetry is a contingent symmetry that depends on the background around which the theory is defined.

It can be observed that in four dimensions, the u-SUSY scheme, NJL terms of the form 
\begin{equation}
\kappa \left[\left(\overline{\psi}_r (T^i)^r{}_s \psi^s\right) \left(\overline{\psi}_p (T^i)^p{}_q \psi^q\right) - \left(\overline{\psi}_r (T^i)^r{}_s\,\gamma_5 \psi^s\right) \left(\overline{\psi}_p (T^i)^p{}_q\,\gamma_5  \psi^q\right)\right] \,,
\end{equation}
are generically produced. They come with completely fixed coefficients as a precise prediction of the model \cite{AVZ2, APZ, AVZ3}. The presence of such terms, where the $SU(N)$ indices $r,s,p,q$ label the different fermionic families, has been proposed as a possible explanation for left-handed neutrino oscillations without a need to assume nonvanishing neutrino masses or sterile features for the right-handed states \cite{Wolfenstein,Fabbri}. Hence, u-SUSY could perhaps provide a mechanism for the missing neutrino puzzle, which could be an indication for an underlying supersymmetry in nature.

We are currently studying the theory defined by \eqref{Lgen} for the superalgebra $su(2,2|2)$ \cite{ADVZ}. Without using the matter ansatz this can be viewed as the mother SUGRA theory of the u-SUSY model \cite{AVZ2}. The Lagrangian describes $u(1)\oplus su(2)$-colored gravitini, with a Rarita-Schwinger Kinetic term, Pauli couplings, NJL terms, interacting with Einstein-Hilbert type of fields and $u(1)\oplus su(2)$ Yang-Mills fields. The supersymmetries are  broken in general, reminding us the Chern-Weil theorem discussed earlier, however the supersymmetry conditions select supergravity backgrounds of a standard fashion. As a consequence, the model \eqref{Lgen} describe an interesting (non-arbitrary) class of supergravity breaking mechanism, which preserves certain isotropy subalgebras. Then applying the matter ansatz, this model produces a spin-$1/2$ model, similar to \cite{AVZ2}, with the same kind of interactions, thus providing Lagrangians of a standard form, including NJL interactions and Pauli terms.

Another extension of the u-SUSY idea to be explored further is the possibility of considering superalgebras, like the one obtained by the supersymmetric completion of two internal gauge symmetries. Consider, for example,  the superalgebra containing $u(1) \oplus su(2) \oplus su(3)$, with generators $\mathbb{K}$, $\mathbb{K}_I$ and $\mathbb{T}_R$ respectively, and supercharges $\mathbb{Q}^\alpha_{ir}$ and $\overline{\mathbb{Q}}_\beta^{js}$. The expected superalgebra would be something like
\begin{align}
[\mathbb{K}_I, \mathbb{K}_J]=i\epsilon_{JK}{}^L\, \mathbb{K}_L\,, \quad [\mathbb{T}_R, \mathbb{T}_S]=if_{RS}{}^U\, \mathbb{T}_U\,, \qquad \qquad \\ \label{QQ'}
\{\mathbb{Q}^\alpha_{ir}\,, \, \overline{\mathbb{Q}}_\beta^{js}\} \sim  \delta^\alpha _\beta \left[\delta^s_r (\sigma^I)^j_i \mathbb{K}_I +  \delta^j_i (\tau^R)^s_r \mathbb{T}_R - i\delta^s_r \delta^j_i \mathbb{K}\right] + \delta^s_r\, \delta^j_i (\Gamma^A)^\alpha{}_\beta  \,\mathbb{J}_A\,.
\end{align}
The connection one-form would be of the form
\begin{equation} \label{A3}
\mathbb{A} =A \mathbb{K} + A^I \; \mathbb{K}_I + A^R \mathbb{T}_R + \omega^A \, \mathbb{J}_A + \overline{\mathbb{Q}}^{ir}_\alpha \, \chi^\alpha_{ir} + \overline{\chi}^{ir}_\alpha \, \mathbb{Q}^\alpha_{ir}\;,
\end{equation}
In this case the fermion fields would be spin 3/2 charged with respect to all three internal gauge charges.

%%%%%%%%%%%%%%%%%%%%%%%%%%%%%%%%%%%%%%%%%%
\section*{Acknowledgments}
%%%%%%%%%%%%%%%%%%%%%%%%%%%%%%%%%%%%%%%%%%
This work has been partially funded by Fondecyt grant 1180368, by MINEDUC-UA project ANT1755, and by Semillero de investigaci\'on project SEM18-02 from Universidad de Antofagasta, Chile. The Centro de Estudios Cient\'{i}ficos (CECs) is funded by the Chilean Government through the Centers of Excellence Base Financing Program of Conicyt
%%%%%%%%%%%%%%%%%%%%%%%%%%%%%%%%%%%%%%%%%%

\end{document}